# The origin of the 'local' ¼ keV X-ray flux in both charge exchange and a hot bubble

M. Galeazzi[1], M. Chiao[2], M. R. Collier[2], T. Cravens[3], D. Koutroumpa[4], K. D. Kuntz[5], R. Lallement[6], S. T. Lepri[7], D. McCammon[8], K. Morgan[8], F. S. Porter[2], I. P. Robertson[3], S. L. Snowden[2], N. E. Thomas[2], Y. Uprety[1], E. Ursino[1], B. M. Walsh[2]

The Solar neighborhood is the closest and most easily studied sample of the Galactic interstellar medium, an understanding of which is essential for models of star formation and galaxy evolution. Observations of an unexpectedly intense diffuse flux of easily-absorbed ¼-keV X rays[1,2], coupled with the discovery that interstellar space within ~100 pc of the Sun is almost completely devoid of cool absorbing gas[3] led to a picture of a "local cavity" filled with X-ray emitting hot gas dubbed the local hot bubble[4-6]. This model was recently upset by suggestions that the emission could instead be produced readily within the solar system by heavy solar wind ions charge exchanging with neutral H and He in interplanetary space[7-11], potentially removing the major piece of evidence for the existence of million-degree gas within the Galactic disk[12-15]. Here we report results showing that the total solar wind charge exchange contribution is 40%±5% (stat) ±5% (sys) of the ¼ keV flux in the Galactic plane. The fact that the measured flux is not dominated by charge exchange supports the notion of a million-degree hot bubble of order 100 pc extent surrounding the Sun.

When the highly ionized solar wind interacts with neutral gas, an electron may hop from a neutral to an outer orbital of an ion, in what is known as charge exchange. The electron then cascades to the ground state of the ion, often emitting soft X rays in the process[16]. The calculations of X-ray intensity from solar wind charge exchange depend on limited information on heavy ion fluxes and even more uncertain atomic cross sections. The DXL sounding rocket mission[17] was launched from White Sands Missile Range in New Mexico on December 12, 2012 to make an empirical measurement of the charge exchange flux by observing a region of higher interplanetary neutral density (with a correspondingly higher charge exchange rate) called the "helium focusing cone". Neutral interstellar gas flows at ~25 km s$^{-1}$ through the solar system due to the motion of the Sun through a small interstellar cloud. This material, mostly hydrogen atoms with about 15% helium, flows from the Galactic direction (l,b)~(3°,16°), placing the Earth downstream of the Sun in early December[18]. The trajectories of the neutral interstellar helium atoms are governed primarily by gravity, executing hyperbolic Keplerian orbits and forming a relatively high-density focusing cone downstream of the Sun about 6° below the ecliptic plane (Fig. 1)[19]. Interstellar hydrogen, on the other hand, also experiences significant impact from radiation pressure and ionization, creating a neutral hydrogen cavity around the Sun, with negligible focusing.

[1]Department of Physics, University of Miami, Coral Gables, FL, 33124, U.S.A. [2]NASA's Goddard Space Flight Center, Greenbelt, MD, 20771, U.S.A. [3]Department of Physics and Astronomy, University of Kansas, Lawrence, KS 66045, U.S.A. [4]Université Versailles St-Quentin; Sorbonne Université, UPMC Univ. Paris 06 & CNRS/INSU, LATMOS-IPSL, 78280 Guyancourt, France. [5]The Henry A. Rowland Department of Physics and Astronomy, Johns Hopkins University, Baltimore, MD 21218, U.S.A. [6]GEPI Observatoire de Paris, CNRS, Université Paris Diderot, 92190, Meudon, France. [7]Department of Atmospheric, Oceanic, and Space Sciences, University of Michigan, Ann Arbor, MI 48109, U.S.A. [8]Department of Physics, University of Wisconsin, Madison, WI 53706, U.S.A.

The early December launch of DXL placed the He focusing cone near the zenith at midnight. The 7° field of view was scanned slowly back and forth across one side of the cone and more rapidly in a full circle to test the consistency of the derived charge exchange contribution in other directions and to make measurements of the detector particle background while Earth-looking (Extended Data Fig. 1). Figure 2 shows the ROSAT All Sky Survey ¼ keV map[20] (R12 band) with the path of the DXL slow scan (red) and fast scan (white) overplotted. The ROSAT observation of the slow-scan region was performed in September 1990 when the line of sight was ~1 A. U. away from, and parallel to, the He cone, so its charge exchange contribution was not significantly affected by the cone enhancement (Fig. 1).

For this analysis, we chose pulse height limits for each DXL proportional counter to most closely match the ROSAT ¼ keV band (Extended Data Fig. 2). This energy range is dominated by and contains most of the "local" emission, whether from solar wind charge exchange or the local hot bubble. To quantify the solar wind charge exchange emission we compared both DXL and ROSAT count rates to well-determined models of the interplanetary neutral distribution along the lines of sight for both sets of measurements (Fig. 3) [17,21]. Figure 4 shows the DXL and ROSAT count rates along the DXL scan path as functions of Galactic longitude. The figure shows the combined Counter-I and Counter-II count rates (black dots) during the DXL scan and ROSAT ¼ keV count rates in the same directions (blue solid line). The best fit to the DXL total count rate (red solid line), and the solar wind charge exchange contributions to DXL (red dashed line) and ROSAT (blue dashed line) rates are also shown (see Table 1 for best-fit parameters: the model shown corresponds to the second column). There is potentially an additional contribution from charge exchange between the solar wind ions and the geocoronal hydrogen surrounding the Earth which tracks the short-term variations in solar wind flux. Its time-variable portion has been removed from the ROSAT maps, and the current best estimate of the residual is about 50 ROSAT Units (RU=$10^{-6}$ counts s$^{-1}$ arcmin$^{-2}$) for the ROSAT ¼ keV band (KDK et al. in preparation). The geocoronal contribution to the DXL flux should be negligible due to the look direction, which is directly away from the sun. The signature of the cone enhancement in the DXL data compared to the ROSAT rates is evident, confirming a significant contribution from charge exchange. However, the best fit shows that the total charge exchange contribution to ROSAT is only about 40%±5% (stat) ±5% (sys) of the total flux observed at the Galactic plane. Its contribution to the ROSAT flux over the DXL scan path is typically ~140 RU. For comparison, the total ROSAT ¼ keV flux ranges from ~300-400 RU in the plane up to 1400 RU in the brightest areas at intermediate and high latitudes. This result implies that the measured fluxes are dominated by interstellar emission, strengthening the original idea of a hot bubble filling the local interstellar medium ~100 pc in all directions.

It has been pointed out that a hot bubble creates an apparent pressure balance problem with the tenuous warm cloud that the Sun is passing through. However, recent results on the magnetic contribution[22] to the cloud pressure and new three-dimensional maps of the local interstellar medium[23] bring the implied pressure of the plasma in the local hot bubble to rough agreement with pressures derived for the local interstellar clouds when the measured contribution from the solar wind charge exchange is removed from the local hot bubble emission (SLS et al., in press on Astrophys. J. Lett.).

## Methods

The total count rate due to charge exchange with H and He is the integral over the line of sight $ds$ of the product of the solar wind ion flux, the donor densities $n_H$ and $n_{He}$, as functions of position, and the sum over products of partial cross sections for producing each X-ray line by charge exchange and the efficiency for producing a detector count from that line:

$$Rate = \int \sum_i \sum_j n_i n_{He} v_{rel} \sigma_i b_{ij} g_j \, ds + \int \sum_i \sum_j n_i n_H v_{rel} \sigma_i b_{ij} g_j \, ds,$$

where $i$ represents the solar wind species, $j$ the emission lines for that species, $\sigma_i$ are the speed-dependent interaction cross sections for individual species, $b_{ij}$ is the line branching ratio, $g_j$ is the instrument's response to line $j$, and $v_{rel}$ is the relative speed between solar wind and neutral flow (both the bulk and thermal velocity). We can then write the ion density $n_i$ in terms of the proton density $n_p$ at $R_o=1$ A.U., assuming that it scales as one over the square of the distance $R$ from the Sun (we verified that neutralization effects on the solar wind ions are negligible), and define the compound cross section as

$$\alpha = \sum_i \sum_j \frac{n_i(R_o)}{n_P(R_o)} \sigma_i b_{ij} g_j.$$

In the case of "constant" solar wind conditions, the solar wind flux can be removed from the integrals, and the total charge exchange rate with H and He can be written as:

$$Rate = n_P(R_o) \, v_{rel} \left( \alpha_{He} \int \frac{n_{He}}{R^2} \, ds + \alpha_H \int \frac{n_H}{R^2} \, ds \right).$$

The assumption of isotropy of the ion flux included in the equation above is an approximation, as the flux is known to vary strongly on time scales ~1 day. Evidence that averaging over the few-week transit time through the relevant interplanetary region smooths these fluctuations is found in the very good agreement between four complete sky surveys performed years apart by different missions[20]. A factor to account for the difference in the solar wind flux between the ROSAT and DXL is included in our fitting procedure.

In this work, we used the expected H and He distribution[21] adapted for solar wind conditions during the respective missions to calculate the integrals for both H and He for each point along the DXL scan path. The distribution of the interplanetary neutrals is calculated based on the solar ionization conditions derived from measurements of backscattered solar radiation and checked by in-situ sampling, so the integrals above can be calculated with some confidence for all lines of sight. We took as free parameters the combination $n_p(R_0) \, v_{rel} \, \alpha_{He}$, the ratio of solar wind fluxes for the two missions $(n_p(R_0)v_{rel})_{DXL}/(n_p(R_0)v_{rel})_{ROSAT}$, and correction factors to fine-tune the calculated ratios of DXL counter responses to ROSAT ¼-keV band response. We then did a global least squares fit for both DXL counter rates for each point along the scan path.

There is insufficient variation in the hydrogen column densities to determine its effective cross section, so for $\alpha_H/\alpha_{He}$ we assumed ratios of one, as assumed in the calculated contributions, and two, since some determinations show smaller cross sections for helium. The residual contribution to ROSAT from charge exchange in the Earth's geocorona has in the past been assumed to be negligible due to zero level agreement with other all-sky surveys done with quite different observing geometries. More recent analyses suggest that the value should be more like

50 RU. Table 1 shows the best-fit parameters for some different assumptions for the residual geocoronal charge exchange contribution to the ROSAT rates and for the ratio of effective cross sections for hydrogen and helium. The total solar wind charge exchange contribution is minimally affected. A systematic error has been included in our results to account for the variation in the table.

There may be other potential sources of systematic uncertainties affecting our result. These include contribution from point sources to the DXL count rate, scattered solar X rays, higher-energy photons into the DXL bands, and a highly non-isotropic solar wind flux. We estimated the first three and found their contribution to our result to be within a few percent, while there is no evidence of a significant non-isotropic solar wind across the DXL slow scan that would significantly affect our result.

**References**

1. Bowyer, C.S., Field, G.B., Mack, J.E.. Detection of an Anisotropic Soft X-ray Background Flux. *Nature* 217, 32-34 (1968).
2. Bunner, A.N., Coleman, P.L., Kraushaar, W.L., McCammon D., Palmieri, T.M., Shilepsky, A., Ulmer, M., Soft X-Ray Background Flux. *Nature* 223, 1222-1226 (1969).
3. Jenkins, E. B., Meloy, D. A., A survey with Copernicus of interstellar O VI absorption. *Astrophys. J.* 193, L121-L125 (1974).
4. Sanders, W.T., Kraushaar, W.L., Nousek J.A., Fried, P. M., Soft diffuse X rays in the southern Galactic hemisphere. *Astrophys. J.* 217, L87-L91 (1977).
5. Cox, D.P., Anderson, P.A., Extended adiabatic blast waves and a model of the soft X-ray background. *Astrophys. J.* 253,268-289 (1982).
6. Snowden, S.L., Cox, D. P., McCammon, D., Sanders, W. T., A model for the distribution of material generating the soft X-ray background. *Astrophys. J.* 354, 211-219 (1990).
7. Cox, D.P., Modeling the Local Bubble. *Lecture Notes in Physics* 506, 121-131 (1998).
8. Cravens, T.E., Heliospheric X-ray Emission Associated with Charge Transfer of the Solar Wind with Interstellar Neutrals. *Astrophys. J.* 532, L153-L156 (2000).
9. Lallement, R., The heliospheric soft X-ray emission pattern during the ROSAT survey: inferences on Local Bubble hot gas. *Astron. Astrophys.*, 418, 143-150 (2004).
10. Welsh, B. Y., Lallement, R., Highly ionized gas in the local ISM: some like it hot? *Astron. Astrophys.* 436, 615-632 (2005).
11. Koutroumpa, D., Lallement, R., Raymond, J. C., Kharchenko, V., The solar wind charge-transfer X-Ray emission in the 1/4 keV energy range: inferences on Local Bubble hot gas at low z. *Astrophys. J.* 696, 1517-1525 (2009).
12. Frisch, P. C., The nearby interstellar medium. *Nature* 293, 377-379 (1981).
13. Cox, D.P., Snowden, S., Perspective on the local interstellar medium. *Adv. Space. Res.* 6, 97-107 (1986).
14. Lallement, R., The local interstellar medium: peculiar or not? *Space Sci. Rev.* 130, 341-353 (2007).
15. Welsh, B. Y., Shelton, R. L., The trouble with the Local Bubble. *Astrophys. Space Sci.* 323, 1-16 (2009).
16. Cravens, T. E., Comet Hyakutake X-ray source: Charge transfer of solar wind heavy ions. *Geophys. Res. Lett.* 24, 105-108 (1997).
17. Galeazzi, M., *et al.*, DXL: a sounding rocket mission for the study of solar wind charge exchange and local hot bubble X-ray emission. *Exp. Astron.* 32, 83-99 (2011).

18. Möbius, E., et al., Synopsis of the interstellar He parameters from combined neutral gas, pickup ion and UV scattering observations and related consequences. *Astron. Astrophys.* 426, 897-907 (2004).
19. Frisch, P. C., The galactic environment of the Sun. *J. Geophys. Res.* 105, 10279-10290 (2000).
20. Snowden, S. L., Freyberg, M. J., Plucinsky, P. P., Schmitt, J. H. M. M., Truemper, J., Voges, W., Edgar, R. J., McCammon, D., Sanders, W. T., First Maps of the Soft X-Ray Diffuse Background from the ROSAT XRT/PSPC All-Sky Survey. Astrophys. J. 454, 643-653 (1995).
21. Koutroumpa, D., Lallement, R., Kharchenko, V., Dalgarno, A., Pepino, R., Izmodenov, V., Quémerais, E., Charge-transfer induced EUV and soft X-ray emissions in the heliosphere. *Astron. Astrophys.* 460, 289-300 (2006).
22. Burlaga, L. F., Ness, N. F., Voyager 1 Observations of the Interstellar Magnetic Field and the Transition from the Heliosheath. *Astrophys. J.* 784, 146 (2014).
23. Puspitarini, L., Lallement, R., Vergely, J. L., Snowden, S. L., Local ISM 3D distribution and soft X-ray background: Inferences on nearby hot gas and the North Polar Spur. *arXiv:1401.6899* (2014).

**Acknowledgments**

We thank the personnel at NASA's Wallops Flight Facility and White Sands Military Range for their support of payload development, integration, and launch, and technical personnel at the University of Miami, NASA's Goddard Space Flight Center, and the University of Michigan for their support of instrument's development. This work was supported by NASA's award #NNX11AF04G. D.K. and R.L. acknowledge financial support for their activity by the program "Soleil Héliosphère Magnétosphère" of the French space agency CNES, and the National Program "Physique Chimie du Milieu Interstellaire" of the Institut national des sciences de l'univers, INSU.

**Author Contribution**

Y.U., N.E.T., M.G., D.M., M.R.C., F.S.P., and S.T.L., contributed to hardware development, Y.U., N.E.T., M.G., D.M., M.R.C., F.S.P., M.C., D.K., K.D.K., and K. M. contributed to launch operations, M.G., D.M., Y.U., N.E.T., M.R.C., D.K., K.D.K., K.M., T.C., I.P.R., D.G.S., S.L.S., E.U., and B.M.W. contributed to data reduction and analysis. D.K. and R.L. prepared the neutral integral distributions. All authors discussed the results and commented on the manuscript.

**Author Information**

Reprint and permissions information is available at www.nature.com/reprints. Correspondence and requests for materials should be addressed to M.G. (galeazzi@physics.miami.edu). B.M.W. current address is Space Sciences Laboratory, University of California, Berkeley, CA 94720 USA. M.C. and N.E.T. are employed through CRESST and the University of Maryland, Baltimore County, MD 21250, USA.

| ROSAT Geocoronal SWCX (RU) | 0 | 50 | 0 | 50 |
|---|---|---|---|---|
| $\alpha_H/\alpha_{He}$ | 1 | 1 | 2 | 2 |
| C-I* | 0.91±0.06 | 0.96±0.06 | 0.87±0.07 | 0.92±0.07 |
| *C-II** | 0.97±0.07 | 1.02±0.07 | 0.92±0.07 | 0.99±0.07 |
| $n_p(R_o)v_{rel}\alpha_{He}$ (RU cm$^3$ A.U.) | 5223±770 | 3197±720 | 4500±490 | 3180±460 |
| DXL/ROSAT Solar wind flux** | 0.63±0.13 | 0.91±0.25 | 0.73±0.14 | 0.96±0.20 |
| $\chi^2$ (136 d.o.f) | 207 | 196 | 206 | 194 |
| Total SWCX contribution to ROSAT R12 data*** | **(39±6)%** | **(37±8)%** | **(39±4)%** | **(42±6)%** |

**Table 1. Best Fit Parameters and maximum solar wind charge exchange contribution to the ¼ keV flux.** Summary of best fit parameters for different assumptions for Geocoronal contribution to ROSAT R12 band, and $a_H/a_{He}$. *C=ratio of fit DXL response to nominal value from laboratory calibrations. The corrections are well within the range expected from spectral uncertainties. **Ratio of solar wind fluxes during the DXL and ROSAT measurements. We note that while both missions were near solar maximum (and therefore should have similar isotropic composition), solar activity in terms of sunspots was weaker during the DXL measurement, as reflected by the fitted ratios. ***Total solar wind charge exchange contribution to ROSAT (interplanetary + geocoronal) divided by observed R12 rate given as percentage of R12 at b=0° (333 RU -- lowest anywhere on scan and close to lowest on sky). Errors are 1-sigma.

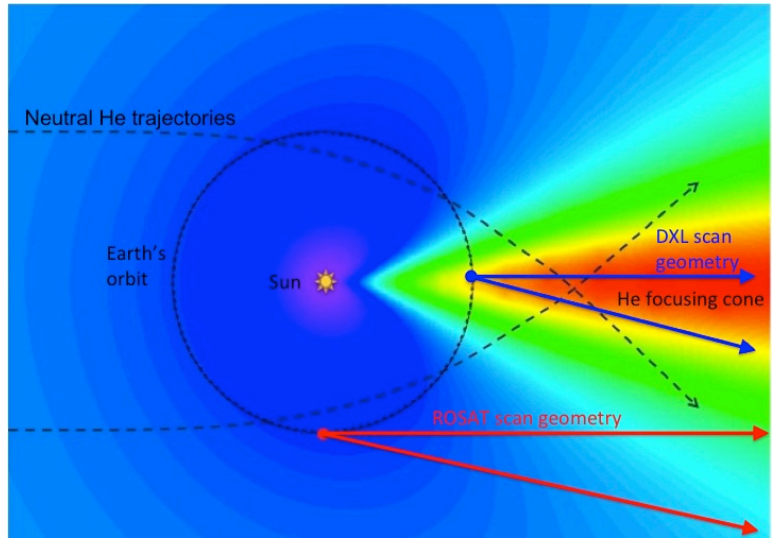

**Fig. 1. The He focusing cone.** Modeled interstellar He density showing the He focusing cone. Keplerian He orbits, the Earth's orbit, and the DXL and ROSAT observing geometries are also shown.

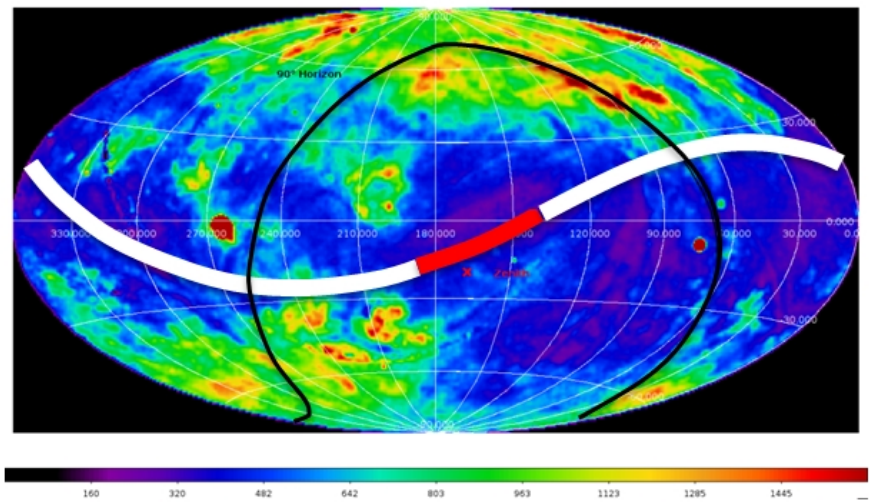

**Fig. 2. The DXL scan path.** ROSAT all-sky survey map in the 1/4-keV (R12) energy band, shown in Galactic coordinates with l=180°, b=0° at the center. Colors show intensity as given by the scale at the bottom. The units are ROSAT Units (RU=$10^{-6}$ cts s$^{-1}$ arcmin$^{-2}$). The DXL scan path is the white band with the slow portion in red. The black line is the 90° horizon for the DXL flight. The width of the band represents the half power diameter of the instrument beam.

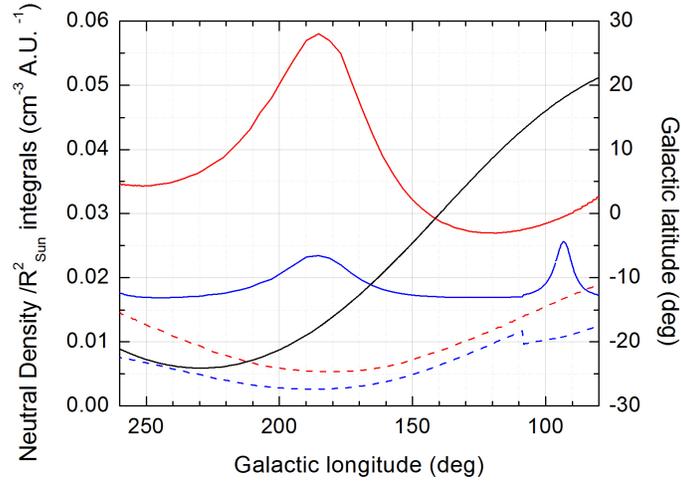

**Fig. 3. Neutral atom column density for DXL and ROSAT.** Weighted neutral column density distribution integrals for each line of sight along the scan path. The red lines are the integral for He (solid) and H (dashed) in the DXL geometry. The blue lines represent the integral for He (solid) and H (dashed) in the ROSAT geometry. The black line shows the galactic latitude during the scan. DXL is significantly more affected by the He focusing cone, while in both cases the H contribution is small.

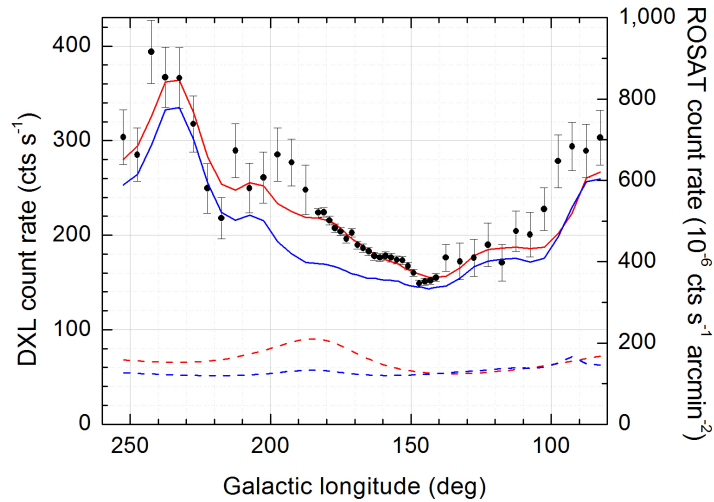

**Fig. 4. Fit to DXL and ROSAT data.** Combined Counter-I and Counter-II count rates (black dots) during the DXL scan and ROSAT ¼ keV count rate in the same directions (blue solid line). The best fit to the DXL total count rate (red solid line), and the solar wind charge exchange contribution to DXL (red dashed line) and ROSAT ¼ keV bands (blue dashed line) are also shown. The error bars are s.e.m.

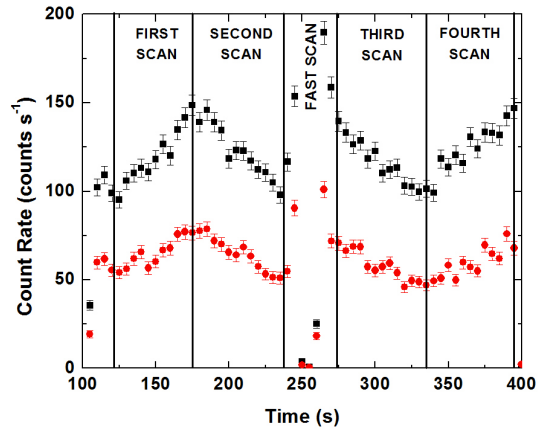

**Extended Data Fig. 1. DXL count rates versus time during flight.** Count rate of Counter-I (red) and Counter-II (black) as a function of time during launch. The DXL observation started off the cone, moved toward the nose of the cone (first scan) and back (second scan) at about 0.7 deg/s, then performed an Earth scan at about 10 deg/s for to measure the instrument background (fast scan), returned to the nose of the cone, and then performed another scan off the cone (third scan) and back to the nose (fourth scan), for a total of 4 slow scans along the cone and 1 fast scan. The gradient up and down the He focusing cone is quite evident. The error bars are s.e.m.

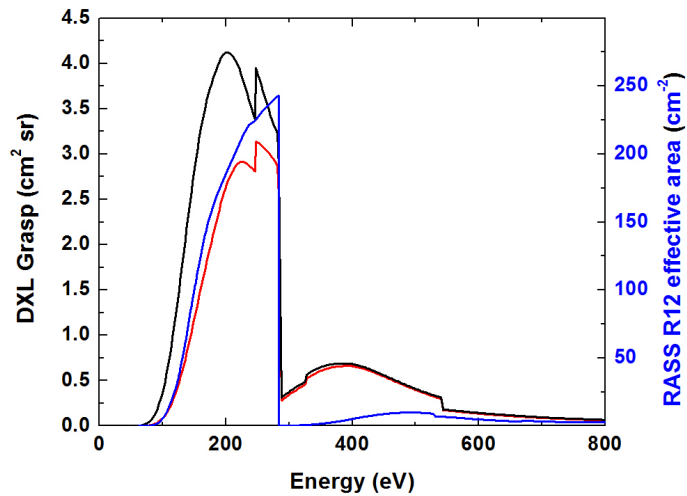

**Extended Data Fig. 2. The DXL detector response.** Effective grasp for the DXL Counter-I (red) and Counter-II (black), compared with the ROSAT ¼ keV (R12 band) effective area (blue).